\documentclass[12pt,a4paper]{article}

\begin{document}

\title{\textbf{A new integrable system of}\\\textbf{symmetrically coupled derivative}\\\textbf{nonlinear Schr\"{o}dinger equations}\\\textbf{via the singularity analysis}}
\author{\textsc{S. Yu. Sakovich\medskip}\\{\small Institute of Physics,}\\{\small National Academy of Sciences,}\\{\small P.O. 72, Minsk, BELARUS.}\\{\small sakovich@dragon.bas-net.by}
\and \textsc{Takayuki Tsuchida\medskip}\\{\small Department of Physics,\vspace{-0.1in}}\\{\small Graduate School of Science,\vspace{-0.1in}}\\{\small University of Tokyo,\vspace{-0.1in}}\\{\small Hongo 7-3-1, Bunkyo-ku,\vspace{-0.1in}}\\{\small Tokyo 113-0033, JAPAN.\vspace{-0.1in}}\\{\small tsuchida@monet.phys.s.u-tokyo.ac.jp\medskip}}
\date{\textsl{April 13, 2000}}
\maketitle
\begin{abstract}
A new integrable system of two symmetrically coupled derivative nonlinear
Schr\"{o}dinger equations is detected by means of the singularity analysis. A
nonlinear transformation is proposed which uncouples the equations of the new system.\bigskip
\end{abstract}

In this paper, we study the integrability of the following system of two
symmetrically coupled derivative nonlinear Schr\"{o}dinger equations:%
\begin{equation}%
\begin{array}
[c]{l}%
q_{t}=\mathrm{i}q_{xx}+aq\bar{q}q_{x}+bq^{2}\bar{q}_{x}+cr\bar{r}q_{x}%
+dq\bar{r}r_{x}+eqr\bar{r}_{x}+\mathrm{i}fq^{2}\bar{q}+\mathrm{i}gqr\bar
{r},\smallskip\\
r_{t}=\mathrm{i}r_{xx}+ar\bar{r}r_{x}+br^{2}\bar{r}_{x}+cq\bar{q}r_{x}%
+dr\bar{q}q_{x}+erq\bar{q}_{x}+\mathrm{i}fr^{2}\bar{r}+\mathrm{i}grq\bar{q},
\end{array}
\label{sys}%
\end{equation}
where $a,b,c,d,e,f,g$ are real parameters, and the bar denotes the complex
conjugation. By means of the singularity analysis, we detect one new
integrable case of the system (\ref{sys}), characterized by the conditions%
\begin{equation}
a=c=e\neq0,\quad b=d=g=0. \label{new}%
\end{equation}
Then we propose a nonlinear transformation, which uncouples the equations
(\ref{sys}) in the case (\ref{new}).

We follow the Weiss-Kruskal algorithm of the singularity analysis \cite{WTC},
\cite{JKM}. With respect to $q,\bar{q},r,\bar{r}$, which should be considered
as mutually independent, the system (\ref{sys}) is a normal system of four
second-order equations, of total order eight. A hypersurface $\phi(x,t)=0$ is
non-characteristic for (\ref{sys}) if $\phi_{x}\neq0$, and we set $\phi_{x}%
=1$. The substitution of the expansions%
\begin{equation}%
\begin{array}
[c]{l}%
q=q_{0}(t)\phi^{\alpha}+\cdots+q_{n}(t)\phi^{n+\alpha}+\cdots,\\
\bar{q}=\bar{q}_{0}(t)\phi^{\beta}+\cdots+\bar{q}_{n}(t)\phi^{n+\beta}%
+\cdots,\\
r=r_{0}(t)\phi^{\gamma}+\cdots+r_{n}(t)\phi^{n+\gamma}+\cdots,\\
\bar{r}=\bar{r}_{0}(t)\phi^{\delta}+\cdots+\bar{r}_{n}(t)\phi^{n+\delta
}+\cdots
\end{array}
\label{exp}%
\end{equation}
(the bar does not mean the complex conjugation now) into the system
(\ref{sys}) determines the branches, i.e.\ admissible choices of $\alpha
,\beta,\gamma,\delta$\ and $q_{0},\bar{q}_{0},r_{0},\bar{r}_{0}$, as well as
the positions $n$ of resonances for those branches. We require that the system
(\ref{sys}) admits a singular generic branch, where at least one of the
exponents $\alpha,\beta,\gamma,\delta$\ is negative, the number of resonances
is eight, and seven of them lie in nonnegative positions. For this singular
generic branch, there are three possibilities: (i) $\alpha+\beta=\gamma
+\delta=-1$, (ii) $\alpha+\beta=-1,\gamma+\delta>-1$, and (iii) $\alpha
+\beta>-1,\gamma+\delta=-1$, which is related to (ii) by $q\leftrightarrow
r,\bar{q}\leftrightarrow\bar{r}$. The choice of (i) would lead us to known
integrable cases of (\ref{sys}), and we will give a detailed consideration of
it in a separate paper. In this paper, we study the possibility (ii), which
leads to a new integrable case of (\ref{sys}). Computations are done by using
the \textit{Mathematica} system \cite{Wol}.

Setting $\beta=-1-\alpha,\gamma+\delta>-1$ in (\ref{exp}), and assuming
without loss of generality that $q_{0}\bar{q}_{0}=\mathrm{i}$, we obtain from
(\ref{sys}) the following four cases:$\medskip$%
\begin{equation}%
\begin{array}
[c]{c}%
a=\frac{-2-5\alpha-5\alpha^{2}}{1+2\alpha},\quad b=\frac{-3\alpha-3\alpha^{2}%
}{1+2\alpha},\medskip\\
c=\frac{(1+2\alpha)e+(1+\alpha)(\gamma-\gamma^{2})-\alpha(\delta-\delta^{2}%
)}{(1+\alpha)\gamma+\alpha\delta},\quad d=\frac{(\alpha\gamma+(1+\alpha
)\delta)e+\gamma\delta(2-\gamma-\delta)}{(1+\alpha)\gamma+\alpha\delta
},\medskip\\
(n+1)n^{3}(n-2)(n-3)(n+\gamma+\delta+\frac{(1+2\alpha)e-(2\alpha+\gamma
)\delta}{(1+\alpha)\gamma+\alpha\delta})\times\medskip\\
(n+\gamma+\delta-2-\frac{(1+2\alpha)e-(2\alpha+\gamma)\delta}{(1+\alpha
)\gamma+\alpha\delta})=0;\medskip
\end{array}
\label{c1}%
\end{equation}%
\begin{equation}%
\begin{array}
[c]{c}%
a=\frac{-2-5\alpha-5\alpha^{2}}{1+2\alpha},\quad b=\frac{-3\alpha-3\alpha^{2}%
}{1+2\alpha},\medskip\\
c=\frac{(1+\alpha)d}{\delta}-\frac{1+2\alpha+2\alpha^{2}}{1+2\alpha}%
+\frac{(1+3\alpha+3\alpha^{2})\delta}{(1+\alpha)(1+2\alpha)},\medskip\\
e=\frac{\alpha\delta(2+2\alpha-\delta)}{(1+\alpha)(1+2\alpha)},\quad
\gamma=-\frac{\alpha\delta}{1+\alpha},\medskip\\
(n+1)n^{3}(n-2)(n-3)(n-\frac{(1+\alpha)d}{\delta}+\frac{2\alpha^{2}}%
{1+2\alpha}+\frac{(1+3\alpha+\alpha^{2})\delta}{(1+\alpha)(1+2\alpha)}%
)\times\medskip\\
(n+\frac{(1+\alpha)d}{\delta}-\frac{2(1+\alpha)^{2}}{1+2\alpha}+\frac
{(1+\alpha-\alpha^{2})\delta}{(1+\alpha)(1+2\alpha)})=0;\medskip
\end{array}
\label{c2}%
\end{equation}%
\begin{equation}%
\begin{array}
[c]{c}%
a=\frac{-2-5\alpha-5\alpha^{2}}{1+2\alpha},\quad b=\frac{-3\alpha-3\alpha^{2}%
}{1+2\alpha},\quad d=e=\gamma=\delta=0,\medskip\\
(n+1)n^{3}(n-2)(n-3)(n-1+c)(n-1-c)=0;\medskip
\end{array}
\label{c3}%
\end{equation}%
\begin{equation}%
\begin{array}
[c]{c}%
a=2,\quad b=e=0,\quad c=\frac{d}{\gamma}+1-\gamma,\quad\alpha=-1,\quad
\delta=0,\medskip\\
(n+1)n^{3}(n-2)(n-3)(n+\frac{d}{\gamma}+\gamma)(n-\frac{d}{\gamma}%
-2+\gamma)=0.\medskip
\end{array}
\label{c4}%
\end{equation}

In the case (\ref{c1}), taking into account that the equation for $n$ should
admit four positive integer solutions and that $\gamma,\delta$ should be
integers, we have to set\medskip%
\begin{equation}
e=\delta-\frac{\delta^{2}}{1+2\alpha},\quad\gamma=-\delta.\medskip\label{n0}%
\end{equation}
Now the positions of resonances are $n=-1,0,0,0,1,1,2,3$. We obtain from
(\ref{sys}) and (\ref{exp}) the recursion relations for $q_{n},\bar{q}%
_{n},r_{n},\bar{r}_{n}$, $n=0,1,\ldots$, and then check the consistency of
those relations at the resonances. At $n=1$, we have to set$\medskip$%
\begin{equation}
g=0,\quad\delta=-1-2\alpha,\quad\alpha(\alpha+1)=0,\medskip\label{n1}%
\end{equation}
otherwise logarithmic terms should be introduced into the expansions
(\ref{exp}). With (\ref{n1}), the recursion relations become consistent at
$n=2,3$ as well. Both choices of $\alpha$, $\alpha=-1$ and $\alpha=0$, lead
through (\ref{n1}), (\ref{n0}), (\ref{c1}) and scaling of variables to the
case (\ref{new}) of the system (\ref{sys}).

In the cases (\ref{c2}) and (\ref{c4}), it is impossible to have four
resonances in positive integer positions, if $\gamma,\delta$ are integers.

In the case (\ref{c3}), we should set $c=0$ to have correct positions of
resonances. Then the consistency of recursion relations at $n=1$ requires
$g=0$, and the system (\ref{sys}) becomes a pair of non-coupled equations.

As we see, the system (\ref{sys}) admits a good singular generic branch in the
new case (\ref{new}). Setting $a=2$ without loss of generality and making the
transformation%
\begin{equation}%
\begin{array}
[c]{c}%
q=q^{\prime}(x^{\prime},t^{\prime})\exp(\mathrm{i}\omega),\;\bar{q}=\bar
{q}^{\prime}(x^{\prime},t^{\prime})\exp(-\mathrm{i}\omega),\smallskip\\
r=r^{\prime}(x^{\prime},t^{\prime})\exp(\mathrm{i}\omega),\;\bar{r}=\bar
{r}^{\prime}(x^{\prime},t^{\prime})\exp(-\mathrm{i}\omega),\smallskip\\
\omega=-\frac{f}{2}x^{\prime}+\frac{f^{2}}{4}t^{\prime},\quad x=x^{\prime
}-ft^{\prime},\quad t=t^{\prime},
\end{array}
\label{trf}%
\end{equation}
we obtain the simplified form of the new case:%
\begin{equation}%
\begin{array}
[c]{l}%
q_{t}=\mathrm{i}q_{xx}+2(q\bar{q}q_{x}+r\bar{r}q_{x}+qr\bar{r}_{x}%
),\smallskip\\
r_{t}=\mathrm{i}r_{xx}+2(r\bar{r}r_{x}+q\bar{q}r_{x}+rq\bar{q}_{x}),
\end{array}
\label{sim}%
\end{equation}
where the prime of the new variables is omitted. Besides the singular generic
branch, already considered, and the Taylor expansions, governed by the
Cauchy-Kovalevskaya theorem, we have to study the following two branches,
admitted by the system (\ref{sim}):%
\begin{equation}%
\begin{array}
[c]{c}%
q_{0}\bar{q}_{0}=-\mathrm{i},\;r_{0}\bar{r}_{0}=-\mathrm{i},\;\alpha
=\gamma=1,\;\beta=\delta=-2,\smallskip\\
(n+1)^{2}n^{2}(n-2)^{2}(n-3)^{2}=0;
\end{array}
\label{b1}%
\end{equation}%
\begin{equation}%
\begin{array}
[c]{c}%
q_{0}\bar{q}_{0}=-\mathrm{i},\;r_{0}\bar{r}_{0}=\mathrm{i},\;\alpha
=\delta=-1,\;\beta=\gamma=0,\smallskip\\
(n+1)^{2}n^{2}(n-2)^{2}(n-3)^{2}=0.
\end{array}
\label{b2}%
\end{equation}
The recursion relations turn out to be consistent at all resonances of the
branches (\ref{b1}) and (\ref{b2}). Consequently, the system (\ref{sim}), or
(\ref{sys}) with (\ref{new}), has passed the Painlev\'{e} test for
integrability well.

Using the \textit{Mathematica} package \textbf{condens.m} \cite{GH}, we can
check that the system (\ref{sim}) has two conservation laws for each rank from
1 to (at least) 4. The fact, that the conservation laws appear by pairs, is
highly suggestive that the equations of (\ref{sim}) can be uncoupled by some
transformation. The first conservation laws of (\ref{sim}) are given by%
\begin{equation}%
\begin{array}
[c]{l}%
(q\bar{q})_{t}=\{\mathrm{i}(q_{x}\bar{q}-q\bar{q}_{x})+(q\bar{q})^{2}%
+2q\bar{q}r\bar{r}\}_{x},\smallskip\\
(r\bar{r})_{t}=\{\mathrm{i}(r_{x}\bar{r}-r\bar{r}_{x})+(r\bar{r})^{2}%
+2r\bar{r}q\bar{q}\}_{x}.
\end{array}
\label{fc}%
\end{equation}
We introduce a new set of dependent variables by%
\begin{equation}
q=u\exp(\mathrm{i}\int_{x_{0}}^{x}v\bar{v}\mathrm{d}x^{\prime}),\quad
r=v\exp(\mathrm{i}\int_{x_{0}}^{x}u\bar{u}\mathrm{d}x^{\prime}).\label{uv}%
\end{equation}
Using the first conservation laws (\ref{fc}) and the definition of new
variables (\ref{uv}), we find that the system (\ref{sim}) is equivalent to two
independent Chen-Lee-Liu equations \cite{CLL}:\medskip%
\begin{equation}
u_{t}=\mathrm{i}u_{xx}+2u\bar{u}u_{x},\quad v_{t}=\mathrm{i}v_{xx}+2v\bar
{v}v_{x}.\medskip\label{ind}%
\end{equation}
Here we have assumed that the dependent variables approach zero as
$x\rightarrow x_{0}$. The transformation (\ref{uv}) of (\ref{sim}) into
(\ref{ind}) proves the integrability of the new case (\ref{new}) of the system
(\ref{sys}).\bigskip

\textbf{Acknowledgments.} The work of S.~Yu.~S. was supported in part by the
Fundamental Research Fund of Belarus, grant $\Phi$98-044. The work of T.~T.
was supported by a JSPS Research Fellowship for Young Scientists.

\end{document}